\documentstyle[aas2pp4]{article}
\begin{document}

\def\Kmin{$K_{\rm min}$}
\def\Mo {$M_{\odot}$}
\def\Mt {$M_{2}$}
\def\MJ {$M_{J}$}
\def\kms {km s$^{-1}$}
\def\ms {m s$^{-1}$}
\def\Ro {$R_{\odot}$}
\def\rl {$R_{L,2}$}

\title{THE MASS DISTRIBUTION OF EXTRASOLAR PLANET-CANDIDATES
AND SPECTROSCOPIC-BINARY LOW-MASS COMPANIONS}

\author{Tsevi Mazeh and Dorit Goldberg}
\affil{School of Physics and Astronomy, Raymond and Beverly Sackler
Faculty of Exact Sciences, Tel Aviv University, Tel Aviv, Israel\\
mazeh@wise7.tau.ac.il; dorit@wise.tau.ac.il}

\author{David W. Latham}
\affil{Harvard-Smithsonian Center for Astrophysics, 60 Garden
Street, Cambridge, MA 02138\\
dlatham@cfa.harvard.edu}



\begin{abstract}

Spectroscopic orbits have been reported for nine unseen companions
orbiting solar-type stars with minimum possible masses in the range 0.5
to 10 Jupiter masses.  We compare the mass distribution of these nine
planet candidates with the distribution of low-mass secondaries in
spectroscopic binaries. Although we still have only a very small number
of systems, the two distributions suggest two distinctive populations.
The transition region between the two populations might be at the range of
10--30 Jupiter masses.

\subjectheadings{binaries: spectroscopic --- planetary systems}
\end{abstract}

\section{INTRODUCTION}

Eight candidates for extrasolar planets have been announced over the
past two years (e.g., Marcy \& Butler 1998).  In each case, very precise
stellar radial-velocity measurements, with a precision of about
10~m~s$^{-1}$ or better, indicated the presence of a low-mass unseen
companion orbiting a nearby solar-type star.  The individual masses of
the eight companions are not known, because the inclination angles of
their orbital planes relative to our line of sight could not have been
measured. The minimum masses for the eight candidates, attained for an
inclination angle of $90^{\circ}$, are in the range 0.5 to 7.4 Jupiter
masses (\MJ). These findings render the eight companions to be giant
planets or at least `planet candidates'.

The detections of these eight companions were announced seven to nine years
after a companion of HD 114762 was discovered (Latham et al. 1989),
based on measurements with a lower precision (Latham 1985).  Mazeh,
Latham, \& Stefanik (1996) have shown that the minimum mass for the
companion of HD~114762 is 9.4 \MJ. Therefore, when considering the
emerging population of planet candidates, HD~114762 should be considered
together with the eight new candidates. Table 1 lists the
minimum mass, period and discovery date of the nine objects.  For random
orbital orientations, the expectation value for sin $i$ is 0.76, so the
actual masses of the nine companions are expected to be in the range of
0.6--12 \MJ.

\begin{deluxetable}{lcrlc}
\tablewidth{0pt}
\tablecaption{The Planet-Candidates}
\tablehead{
\colhead{Name} &\colhead{$M_2 \sin i$} &
\colhead{P\phantom{.0}} &
\colhead{Discovery} & \colhead{Ref.}\nl
 & \colhead{$(M_{J})$} & \colhead{(days)}&\colhead{Date}  \nl}
\startdata
HD 114762      &   9.4    &  84{\phantom{.0}}    &  1989  & 1,2 \nl
51 Peg         &   0.5    &  4.2                 &  1995  & 3 \nl
47 UMa         &   2.5    &  1090{\phantom{.0}}  &  1996  & 4 \nl
70 Vir         &   7.4    &  117{\phantom{.0}}   &  1996  & 5 \nl
55 Cnc         &   0.8    &  14.7                &  1996  & 6 \nl
$\tau$ Boo     &   3.9    &  3.3                 &  1996  & 6 \nl
$\upsilon$ And &   0.7    &  4.6                 &  1996  & 6 \nl
16 Cyg B       &   1.6    &  804{\phantom{.0}}   &  1996  & 7 \nl
$\rho$ CrB &   1.1    &   39.6               &  1997  & 8 \nl
\enddata
\tablecomments{$^1$Latham et al. 1989;
$^2$Mazeh, Latham \& Stefanik 1996;
$^3$Mayor \& Queloz 1995;
$^4$Butler \& Marcy 1996;
$^5$Marcy \& Butler 1996;
$^6$Butler et al. 1997;
$^7$Cochran et al. 1997;
$^8$Noyes et al. 1997.}
\end{deluxetable}

The nature of the newly discovered low-mass companions is not yet clear.
They could be planets, as suggested by various authors (e.g., Marcy \&
Butler 1998), or many could just be brown-dwarf secondaries, formed in
binary stars (Black 1997).  With the small, but not insignificant number
of spectroscopic orbits implying planetary minimum masses, we can now
begin to study the distribution of their orbital parameters, in order to
address this very basic question.

In this paper we discuss the emerging difference between the {\it mass}
distribution of planet candidates and the low-mass end of the
distribution of binary secondaries.  This point has been already
discussed by previous studies (Basri \& Marcy 1997; Mayor, Queloz \&
Udry 1998; Mayor, Udry \& Queloz 1998; Marcy \& Butler 1998), but in
those papers the mass distribution was binned linearly. Here we choose
to use a logarithmic scale to study the mass distribution, because of
the large range of masses, 0.5--300 \MJ, involved.  The logarithmic
scale has also been used by Tokovinin (1992) to study the secondary mass
distribution in spectroscopic binaries, and was suggested by Black (1998)
to study the mass distribution of the planetary-mass companions.

This work is based on an extremely small sample, and the validity of our
results will need to be verified by many more detections. However, if
verified, the difference in mass distributions that we find might
provide an important clue for how to distinguish between planets and
low-mass stellar companions.

A preliminary version of this work was presented at the meeting
``Physical Processes in Astrophysical Fluids'', in Haifa, January 1998
(Mazeh 1998).

\section{COMBINED MASS DISTRIBUTION}

We wish to consider the mass distribution of the planet candidates and
compare it with that of the low-mass secondaries in spectroscopic
binaries. To do so we use the results of two very large
radial-velocity studies of spectroscopic binaries recently completed,
for which partial results have been published. One sample is composed
of G and K stars studied by Mayor et al. (1997), and the other is the
Carney \& Latham (1987, hereafter C-L) high-proper-motion sample.  We
will use the first sample to estimate the mass distribution in the
mass range $1\leq log(M/M_J)\leq 2$, and the latter sample to
estimate the value of the distribution in the range $2\leq
log(M/M_J)\leq 2.5$.

Mayor et al. (1997) listed 10 spectroscopic binaries with
minimum secondary masses in the range 10--63 \MJ. Their list 
includes only two systems in the range
 $1 \leq log(M/M_J) \leq 1.5$ and 8 systems in the range 
 $1.5 \leq log(M/M_J) \leq 2$.
We note that their list does not cover
the second range completely, because the table of Mayor et al. does not
include binaries with minimum secondary masses between 63 and 100 \MJ.
Mayor et al. were kind enough to let us know that they have found 5
additional binaries in the range of 
63--100 \MJ\ (Halbwachs, private communication).

Detailed results for the C-L sample are not yet published, but partial
results were presented in two conference papers (Latham et al. 1998;
Mazeh, Goldberg and Latham 1998). Mazeh, Goldberg and Latham divided
the sample into two subsamples, with high- and low-mass primaries. We
use here only the high-mass primary subsample, with primary masses
between 0.7 and 0.85 \Mo, because they are more similar to the primaries in
the other samples considered here. From Figure 1 of Mazeh, Goldberg
and Latham we can estimate the number of systems with secondary masses
in the range $2\leq log(M/M_J) \leq 2.5$ to be 20.

\begin{figure}
\figurenum{1}
\plotone{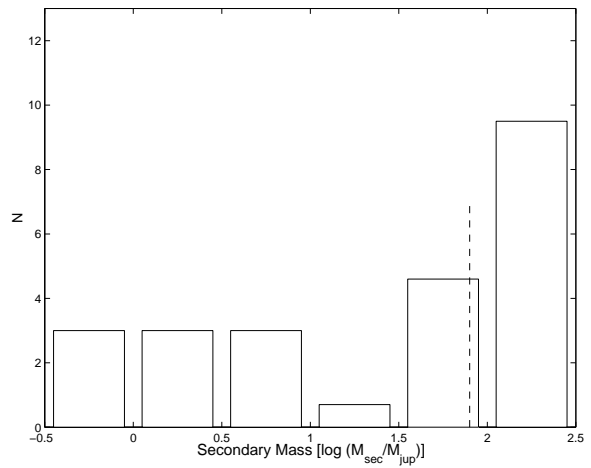}
\caption{Scaled histogram of the extrasolar planet-candidates and the
low-mass secondaries of spectroscopic binaries. The dashed line is the
stellar/substellar limit.}
\end{figure}

The results of the two samples of spectroscopic binaries have to be
scaled to the size of the sample out of which the nine planet candidates
were found.  The scaling is not simple because the parent samples in
which the planets were found are not well defined. The nine planets were
discovered by different research groups, with different time coverage
and slightly different precision (e.g., Marcy \& Butler 1998).  For the
present discussion we will {\it assume} that the total number of stars
searched was two hundred. This has to be compared to the 570 stars of
Mayor et al. (1997) and 420 stars of the sub-sample of high-mass
primaries of the C-L sample (Mazeh, Goldberg \& Latham 1998). The
results are summarized in Table 2, where $N_{scl}$ is the number of
binaries, scaled to a sample of 200 systems. The combined scaled
histogram is plotted in Figure 1.

\begin{deluxetable}{cccr}
\tablewidth{0pt}
\tablecaption{The Scaled Mass Distribution}
\tablehead{\colhead{Mass Range} &\colhead{\# of Observed Systems} &
\colhead{Scaling} &\colhead{$N_{scl}$}\nl
(\MJ)& &\colhead{Factor} &  \nl}
\startdata
$-0.5         \leq log(M) \leq 0.\phn$    & 3          &         &$3\phd\phn$\nl
\phs$0.\phn   \leq log(M) \leq 0.5$       & 3          &         &$3\phd\phn$\nl
\phs$0.5      \leq log(M) \leq 1.\phn$    & 3          &         &$3\phd\phn$\nl
\phs$1.\phn   \leq log(M) \leq 1.5$       & 2          & 200/570 &$0.7      $\nl
\phs$1.5      \leq log(M) \leq 2.\phn$    &$8+5^{\dag}$& 200/570 &$4.6$\nl
\phs$2.\phn   \leq log(M) \leq 2.5$       & 20         & 200/420 &$9.5     $\nl
\enddata
\tablecomments{$^{\dag}$ unpublished data (Halbwachs, private communication)}
\end{deluxetable}

\section{ESTIMATE OF THE CORRECTED DISTRIBUTION}

Before considering the possible interpretation of the combined
histogram, we have to correct the histogram for two effects. The first
one has to do with the fact that the masses given in Table 1 and in
Mayor et al. (1997) list are only {\it minimum} masses, and therefore
the actual mass of each secondary is most probably larger. The
correction of this effect tends to shift the distribution towards the
right side of Figure 1. The second effect reflects the fact that
binaries with too small amplitudes could not have been detected, because
their period is too large, or their inclination angle is too small. The
correction of this effect tends to increase the number of companions
detected in bins with small masses, while the effect is negligible for
bins with large masses. Both effects were taken into account in the work
of Mazeh, Goldberg \& Latham (1998), so we need to correct only the
counts of the two other samples.

To correct for the first effect we calculated the probability of each
system to fall in each bin of the histogram, assuming random orientation
in space. To derive a modified histogram we added up the contributions
of each binary to each bin of the histogram, denoting the resulting
counts by $N_{mod}$.

To correct for the second effect we consider the probability of {\it
not} detecting a binary or a planet in a systematic radial-velocity
search (see Mazeh, Latham \& Stefanik (1996) for details).  Suppose that
the search detects all stars with radial-velocity modulation with a
period $P$ between $P_{\rm min}$ and $P_{\rm max}$, and with
semi-amplitude $K$ larger than or equal to the search threshold $K_{\rm
min}$.  For given primary and secondary masses, and for a given orbital
period, all systems with an inclination smaller than some threshold
inclination {\it cannot} be detected, because $K$ is smaller than
$K_{\rm min}$.  We can therefore calculate $U[K_{\rm min}](P,M_1,M_2)$
--- the probability of {\it not} detecting a binary by a search with a
given threshold $K_{\rm min}$, assuming random orientation in space
(e.g., Mazeh \& Goldberg 1992).

To get the probability of {\it not} detecting a binary taken at random
from a population of binaries, with given {\it range} of secondary masses and
periods, we have to integrate $U[K_{\rm min}](P,M_1,M_2)$ over the given
parameters of the population.  We will then get $U[K_{\rm min}](M_1)$,
which presents the probability of not detecting a
binary, averaged over the secondary masses and period domains.

To correct for the undetected binaries we have to multiply the number
of systems in each bin by the corresponding
\begin{equation}
C = \Bigl(1-U[K_{\rm min}]\Bigr)^{-1} \ , 
\end{equation}
where we have dropped the dependence on $M_1$.

 The main parameter here is \Kmin, which in turn strongly depends on the
precision per measurement, but also on the number of measurements per
star and their temporal distribution. Therefore, the exact values of
\Kmin\ for each of the samples discussed here are still not well
known. For the planet searches we will assume that \Kmin\ is twice the
precision per measurement. This means that any binary with a
peak-to-peak variation larger than 2\Kmin, or four times larger than the
error per measurement of the survey, was detected.  We therefore assume
\Kmin\ to be 20 m s$^{-1}$ for the planet search. For the Mayor et
al. sample the detection threshold is larger than twice the precision
per measurements (Halbwachs, private communication), so we will assume,
somewhat arbitrarily, \Kmin\ of 1 \kms.

To calculate the correction factor for each bin of the histogram, we
considered a population of binaries with secondary mass range coinciding
with the bin mass range, with a Duquennoy \& Mayor (1991) period
distribution between 1 and 1500 days.  We then applied the correction
derived to the modified counts in each bin to get $N_{cor}$.  We
estimated the error in each of the first three bins by the square root
of the modified number of systems in each bin, multiplied by the
correction factor.  For the Mayor et al. sample we took into account the
fact that the original sample was larger by $570/200$, so the {\it
relative} errors for these two bins are smaller by the square root of
this scaling factor. The `corrected' histogram is given in Table 3.

\begin{deluxetable}{cccc}
\tablecolumns{4}
\tablewidth{0pt}
\tablecaption{Estimated Corrected Mass Distribution}
\tablehead{\colhead{Mass Range} &\colhead{$N_{mod}$} &
\colhead{Correction} &\colhead{$N_{cor}$}\nl
\colhead{(\MJ)}&&\colhead{Factor} &  \nl
\cline{1-4}\nl
\multicolumn{4}{c}{Planet Search}}
\startdata
$-0.5         \leq log(M) \leq 0.\phn   $ & 1.6 & 2.2 &$ 3.6\pm 2.8$\nl
\phs$0.\phn   \leq log(M) \leq 0.5      $ & 2.7 & 1.1 &$ 2.9\pm 1.8$\nl
\phs$0.5      \leq log(M) \leq 1.\phn   $ & 2.6 &$1\phd\phn$&$ 2.6\pm
1.6$\nl
\phs$1.\phn   \leq log(M) \leq 1.5      $ & 1.5 &$1\phd\phn$&$ 1.5\pm
1.2$\nl
\cutinhead{Mayor et al. Sample}
\phs$1.\phn   \leq log(M) \leq 1.5      $ & 0.4 & 3.9 &$ 1.6\pm 1.5$\nl
\phs$1.5      \leq log(M) \leq 2.\phn   $ & 3.1 & 1.3 &$ 4.0\pm 1.4$\nl
\cutinhead{Carney \& Latham Sample}
\phs$2.\phn   \leq log(M) \leq 2.5      $ &     &     &$ 9.5\pm 2.1$\nl
\enddata
\end{deluxetable}

Note that both samples cover the range $1. \leq log(M/M_J) \leq 1.5$,
both of which yielded very similar estimates.  When plotting the
corrected histogram in Figure 2 we combined the two estimates together
and got $1.5\pm1.0$ for this bin.

\begin{figure}
\figurenum{2}
\plotone{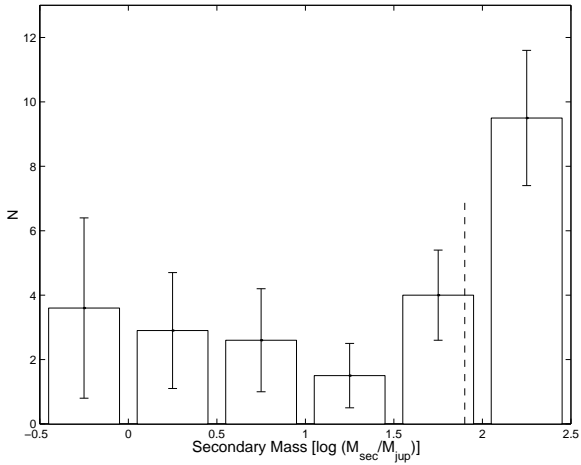}
\figcaption{Corrected histogram of the extrasolar planet-candidates and
the low-mass secondaries of spectroscopic binaries. The dashed line is
the stellar/substellar limit.}
\end{figure}

\section{DISCUSSION}

The corrected combined histogram might suggest that we see here two
populations.  At the high-mass end of the histogram we see a
distribution which drops steeply when we move from 200 to 20 \MJ. At the
planetary range of masses we see a flat distribution, which might even rise
very mildly when we move from, say, 20 to 0.6 \MJ.  Unfortunately, the
number of systems in each bin is small. However, the two different
slopes in the two parts of the diagram seem real, as they are based on
more than one bin.

The derived diagram depends on two parameters --- \Kmin\ and the number
of bins of the histogram. We got the same gross features, namely two
opposite slopes in the two parts of the diagram, when we changed the
values of these two parameters. Dividing the total range of the diagram
into 5 or 4 bins instead of 6 bins shifted the transition region between the
two slopes somewhat to the right. Changing \Kmin\ from 20 to 50 \ms\ made the
slope at the left hand side of the diagram steeper. We conclude
therefore that the overall shape of the diagram does not depend strongly
on the specific values of the parameters of the derivation.

The transition region between the two populations is at about
10--30~\MJ. Unfortunately, the relative error of this bin is very large.
Nevertheless, it seems that this is the bin with the smallest number of
systems. The very low count estimate in this bin is supported by the
fact that the very sensitive searches for planets, which yielded the
discovery of the eight new planet candidates, did not find any
companions with minimum masses between 10 and 30 \MJ.  With \Kmin\ of
20~m~s$^{-1}$ these searches could detect more than 99\% of the binaries
in this bin. The fact that no precise search discovered any binary in
this bin indicates that the number of systems with secondary masses
between 10 and 30 \MJ\ is very small.

The drop of the secondary mass distribution we find when moving from
200 to 20~\MJ\ is consistent with the finding of Halbwachs, Mayor \&
Udry (1998), who studied the mass {\it ratio} distribution of
spectroscopic binaries in the samples of G and K stars of Mayor et
al. (1997).  Halbwachs, Mayor \& Udry found a flat histogram of the mass
ratio, although they could not exclude increasing or decreasing power
laws of the form
$q^{\alpha}$, where $q$ is the mass ratio and $-0.82\leq \alpha\leq
0.87$. The flat distribution of $q$ yields constant $dN/dm_2$, if
all primary masses are similar.  This corresponds to
$dN/d\log(m_2)\propto m_2$, consistent with our findings.
 The drop we find is also consistent with the findings of Mayor, Queloz
\& Udry (1998; see also Mayor, Udry \& Queloz 1998) who found that
$dN/dm_2\propto m_2^{-0.4}$. Their result corresponds to
$dN/d\log(m_2)\propto m_2^{+0.6}$. Figure 2 of this work suggests a steeper
drop, but the difference is within the errors.

The transition region between the two populations, or between the two
slopes, that we find here is, however, different from the findings of
Mayor, Queloz \& Udry (1998) and Mayor, Udry \& Queloz (1998). They find
a borderline at 7 \MJ, while our logarithmic treatment of the data
suggests a transition region at the range of 10--30 \MJ. Another
difference is the shape of the distribution in the planetary mass
range. They find a very steep rising distribution when moving down
towards the range of 1--5 \MJ.  We find an almost flat logarithmic
distribution, with perhaps a mild rise towards lower masses, depending
on the exact value of \Kmin.

Let us {\it assume} that Figure 2 indeed shows two distinctive slopes in
the two parts of the diagram.  Let us further {\it assume} that this
reflects the fact that we see here two different populations, one below
10--30 \MJ, and one with masses larger than this transition region.  One
possible interpretation of the diagram, if indeed we see here two
different populations, is that the two populations were formed
differently. Maybe the lower-mass population was formed like planets,
out of an accretion disc, while the higher-mass population was formed
like binary stars, in a mechanism which probably involves large-scale
gravitational collapse (e.g. Boss 1996; Black 1986).  If this is the
case then the binary secondaries include stars and brown dwarfs
together.

Figure 2 suggests that the transition region between the two populations is at
about 10--30~\MJ. This is of astrophysical significance, if indeed the
lower-mass population is composed of planets, as it might tell us about
the lower limit and upper limit of the formation of secondaries and
planets, respectively (Marcy \& Butler 1995, 1998; Mayor, Queloz \& Udry
1998; Mayor, Udry \& Queloz 1998).  The upper limit of the planetary masses
is set by the conditions in the accretion disc, and most probably by the
interaction between the planet and the gas and dust in the disc. Boss
(1996) already noted that Lin and Papaloizou (1980) theoretically
predicted that the maximum mass for the formation of a planet in the
disc of the Solar nebula is about 1 \MJ. As the maximum mass depends on
the mass of the early nebula, we can get somewhat higher masses in
different cases.  The lower limit for secondary masses in binaries is
set by the binary formation mechanism, whatever that mechanism might be.
Boss (1988), for example, noted that the theory of opacity limited cloud
fragmentation predicts that the minimum mass for a companion is about 10
\MJ.  In fact, Low \& Lynden-Bell (1976) estimated already 20 years ago
that the minimum Jeans mass for fragmentation of a molecular cloud is 7
\MJ. Silk (1977), in a contemporaneous study, came up with minimum
masses between 10 and 100 \MJ, depending on the shape of the
collapse. If we indeed see the transition region between the two populations at
about 10--30 \MJ, this is not too far from the predictions of the
theories.

Duquennoy \& Mayor (1991; see also Mayor, Queloz \& Udry 1998) have
suggested that the observed orbital eccentricities can be used to
distinguish between planets and stellar companions. However, Mazeh,
Mayor \& Latham (1996), when discussing the eccentricity versus mass of
the known planet candidates, pointed out that the planet-disc
interaction (e.g., Goldreich \& Tremaine 1980) is a possible mechanism
for generating a strong dependence of eccentricity versus mass
(Artymowicz 1992; Lubow \& Artymowicz 1996), at least for moderate
eccentricities. This possibility can undermine the potential of the
eccentricity-mass dependence to distinguish between planets and
secondaries. Furthermore, Black (1997) analyzed the eccentricity as a
function of {\it period} and concluded that the eccentricities observed
are consistent with the assumption that all the planet candidates are
actually low-mass brown dwarfs formed like binary stars. It seems
therefore that it might be too early to distinguish between brown dwarfs
and planets solely on the basis of their orbital eccentricity.

Mazeh, Mayor \& Latham (1996) speculated that ``The 10--40 \MJ\ mass gap
may prove to be critical for the interpretation of'' the
eccentricity-mass dependence.  We confirm here that the transition region
between the two populations could be at this range of masses.

Obviously, the left hand side of the histogram and the transition region
between the two slopes derived in this paper are based on a very small
number of objects all together, and these features need to be verified
by many more detections. Further, one still needs to make sure that the
different slope in the planetary-mass range is not due to some selection
effects. For example, there might be a correlation between the orbital
period and the secondary mass, which might make the small-mass
secondaries easier to detect. Such an effect could cause the histogram
to appear to rise towards smaller mass. However, if the shape of the
histogram can be verified, and if the planetary-mass objects prove to be
extrasolar planets, the shape of the histogram might give us the
long-sought clue for how to distinguish planets from low-mass stellar
companions.

We express our thanks to J.-L. Halbwachs for his very useful comments on
the manuscript. We thank M. Mayor, S. Udry and J.-L. Halbwachs for
letting us use their unpublished results.  We thank the referee,
Dr. D. Black, for a critical reading of the manuscript and for his
comments that led to significant improvement of the paper.  This work
was supported by US-Israel Binational Science Foundation grant 94-00284
and by the Israeli Science Foundation.




\end{document}